\documentclass[aps,prl,reprint,nofootinbib]{revtex4-1}
\usepackage{graphicx}	
\usepackage{amsmath}	
\usepackage{amssymb}	
\usepackage{float}
\usepackage{aas_macros}
\usepackage{bm}		
\usepackage{float}
\usepackage{physics}
\usepackage{hyperref}
\usepackage{soul}
\usepackage[capitalise,nameinlink,noabbrev]{cleveref}
\usepackage{xcolor}
\usepackage[caption=false]{subfig}
\usepackage{cancel}
\usepackage{soul}
\usepackage{xcolor}
\newcommand{\D}{\mathrm{d}}

\newcommand\approxinbetween[4]{#2\,\mathrm{#4}\,\lesssim #1\lesssim {#3}\,\mathrm{#4}} 
\newcommand\bv{Brunt-Väisälä frequency }

\newcommand\omegabv{\omega_\mathrm{BV}^2}

\begin{document}
\title{Gravitational Waves from a Core g-Mode in Supernovae as Probes of the High-Density Equation of State}
\author{Pia Jakobus}
\email{pia.jakobus@monash.edu}
\author{Bernhard M\"uller}
\author{Alexander Heger}
\affiliation{School of Physics and Astronomy, Monash University, Clayton, VIC~3800 Australia}
\author{Shuai Zha}
\affiliation{Tsung-Dao Lee Institute, Shanghai Jiao Tong University, Shanghai 200240, China}
\author{Jade Powell}
\affiliation{Centre for Astrophysics and Supercomputing, Swinburne University of Technology, Hawthorn, VIC 3122, Australia}
\author{Anton Motornenko}
\author{Jan Steinheimer}
\author{Horst St\"ocker}
\affiliation{Frankfurt Institute for Advanced Studies, Giersch Science Center, Frankfurt am Main, Germany}

\begin{abstract}
Using relativistic supernova simulations of massive progenitor stars with a quark-hadron equation of state (EoS) and a purely hadronic EoS, we identify a distinctive feature in the gravitational-wave signal that originates from a buoyancy-driven mode (g-mode) below the proto-neutron star convection zone. The mode frequency lies in the range $200\lesssim f\lesssim 800\,\text{Hz}$ and decreases with time. As the mode lives in the core of the proto-neutron star, its frequency and power are highly sensitive to the EoS, in particular the sound speed around twice saturation density.

\end{abstract}
\maketitle
\textit{Introduction.}---Core-collapse supernovae (CCSNe) are among the most important astronomical events yet to be detected by ground-based gravitational-wave (GW) interferometers~\citep{aLigo_2015,adVirgo_2015,kagra_2022,abbott_20,kalogera_19}. 
With current detector sensitivity, the event must occur within at most a few $10\, \mathrm{kpc}$ of Earth~\citep{hayama_2015,abbott_20,detectability_2021}.  Future detectors, such as the Einstein Telescope, may observe supernovae throughout the Milky Way and beyond the Magellanic Clouds \citep{Punturo_2010,gw_review_2030s,powell_2019,srivastava_19}.
The estimated rate for galactic CCSNe is $3^{+7.3}_{-2.6}$ per century \citep{Li2011,Adams2013}, implying a realistic chance of detection within the lifetime of second- and third-generation instruments. Such a detection 
would reveal insights into the properties of the proto-neutron star and the multi-dimensional fluid flow in the supernova core. 
Multi-dimensional simulations \citep[e.g.,][]{Murphy_2009,mueller_2013,kuroda_14,Kuroda_2016,andresen_2017,Yakunin_2017,Takiwaki_2018,Morozova_2018,Andresen_2019,Radice_2019,Andresen_2019,Mezzacappa_2020,vartanyan_23} show that the GW signal reflects the presence of proto-neutron star (PNS) oscillation modes triggered
by convection, turbulent accretion, the standing-accretion shock instability (SASI) \citep{blondin_2003}, or triaxial instabilities. 
The most robust feature in the signal
comes from a quadrupolar $f$/$g$-mode \citep{mueller_2013,sotani_2017,Morozova_2018,Torres_2018,torres_2019,Andersen_2021,mezzacappa_22} with a frequency that increases in time from a few hundred Hz to above $1\, \mathrm{kHz}$. 
Future GW observations may measure this frequency \citep{powell_2022,afle_23}
and use mode relations \citep{mueller_2013,torres_2019b,sotani_21} to constrain bulk PNS parameters (mass, radius, surface temperature).

Unfortunately, the dominant $f$/$g$-mode is largely confined to the PNS surface region and therefore only indirectly sensitive (through the PNS radius) to the high-density equation of state (EoS). 
Nevertheless, potential GW diagnostics that could also constrain the properties of nuclear matter at
several times nuclear saturation density $\rho_0 \geq 2.6\times 10^{14}\,\mathrm{g}\,\mathrm{cm}^{-3}$ and temperatures of several 10\,GK are being identified. The GW signal could shed light on the nuclear EoS, e.g., about the possible appearance of quarks at high densities~\citep{Gentile_1993,Bednarek_1996}, which is not considered in most standard CCSN simulations.  A first-order phase transition from hadrons to quarks~\citep{Sagert_2009,Fischer2017}, which is already known to strongly affect  post-merger GW emission in neutron star mergers \cite{Dexheimer_2018,Most_2019a,Bauswein_2019,Hanauske_2019},
produces a loud and distinct supernova GW signature with a peak at several kHz, regardless of whether the phase transition triggers an explosion \cite{osti_2007,Zha2020,Kuroda2021}.

Here we compare the predicted GW
signals from supernova simulations with
the purely hadronic \texttt{SFHx} EoS
\cite{Steiner_2013}
and the chiral mean field (\texttt{CMF}) EoS \cite{motornenko_2020} with a \emph{smooth crossover} to quark matter. We find that the 
\texttt{CMF} models exhibit a \emph{core} $g$-mode signature of lower frequency 
and higher intensity as a distinct GW fingerprint,
and elucidate the underlying EoS properties.

\textit{Methods.}---We perform axisymetric (2D) simulations with the general-relativistic neutrino hydrodynamics code
\textsc{CoCoNuT-FMT} \cite{mueller_2010,mueller_2015}.
Different from recent multi-D simulations with \textsc{CoCoNuT-FMT}, we calculate only a small inner region of radius $<1\mathord 380\,\text{m}$ in spherical symmetry to capture $g$-modes in the PNS core. GW signals are calculated using a modified version \citep{mueller_2013} of the time-integrated quadrupole formula \citep{finn_1989}. 
We use two zero-metallicity progenitors 
of
$35\,\mathrm{M}_\odot$ and $85\,\mathrm{M}_\odot$ (named
\texttt{z35} and \texttt{z85}), which are calculated with the stellar evolution code \textsc{Kepler}~\cite{weaver_1978,heger_2010}.

We employ two different high-density EoS.
For models \texttt{z35:CMF} and
\texttt{z85:CMF}, we use
the \texttt{CMF} model with a first-order nuclear liquid-vapor phase transition at densities $\mathord{\sim} \rho_0$, a second, but weak first-order phase transition due to chiral symmetry restoration at  $\mathord{\sim}4\mathord{\times}\rho_0$ with a critical endpoint at $T_\text{CeP}\approx 15$~MeV,
and a smooth transition to quark matter at higher densities. The \texttt{CMF} EoS has a ground state density (for symmetric matter) $n_\text{sat}=0.16\,\text{fm}^{-3}$, binding energy per baryon $E_0/B = -15.2\,\text{MeV}$, asymmetry energy $S_0=31.9\,\text{MeV}$, incompressibility $K_0=267\,\text{MeV}$, and a maximum Tolman-Oppenheimer-Volkoff mass $M_\text{TOV}^\text{max}= 2.10\,\mathrm{M}_\odot$~\citep{motornenko_2020}. This EoS has recently been studied in the context of neutron star merger and 1D CCSN simulations~\citep{most_2022,jakobus_2022}. The second EoS, used for runs \texttt{z35:SFHx} and
\texttt{z85:SFHx},
is the purely hadronic relativistic mean-field \texttt{SFHx} model~\cite{Steiner_2013}. Nuclear matter properties for the \texttt{SFHx} EoS are:  $n_\text{sat}=0.16\,\text{fm}^{-3}$, $E_0/B = -16.16\,\text{MeV}$, $S_0=28.67\,\text{MeV}$, $K_0=239\,\text{MeV}$, and $M_\text{TOV}^\text{max}= 2.13\,\mathrm{M}_\odot$.

\begin{figure}
	\includegraphics[width=1.07\columnwidth]{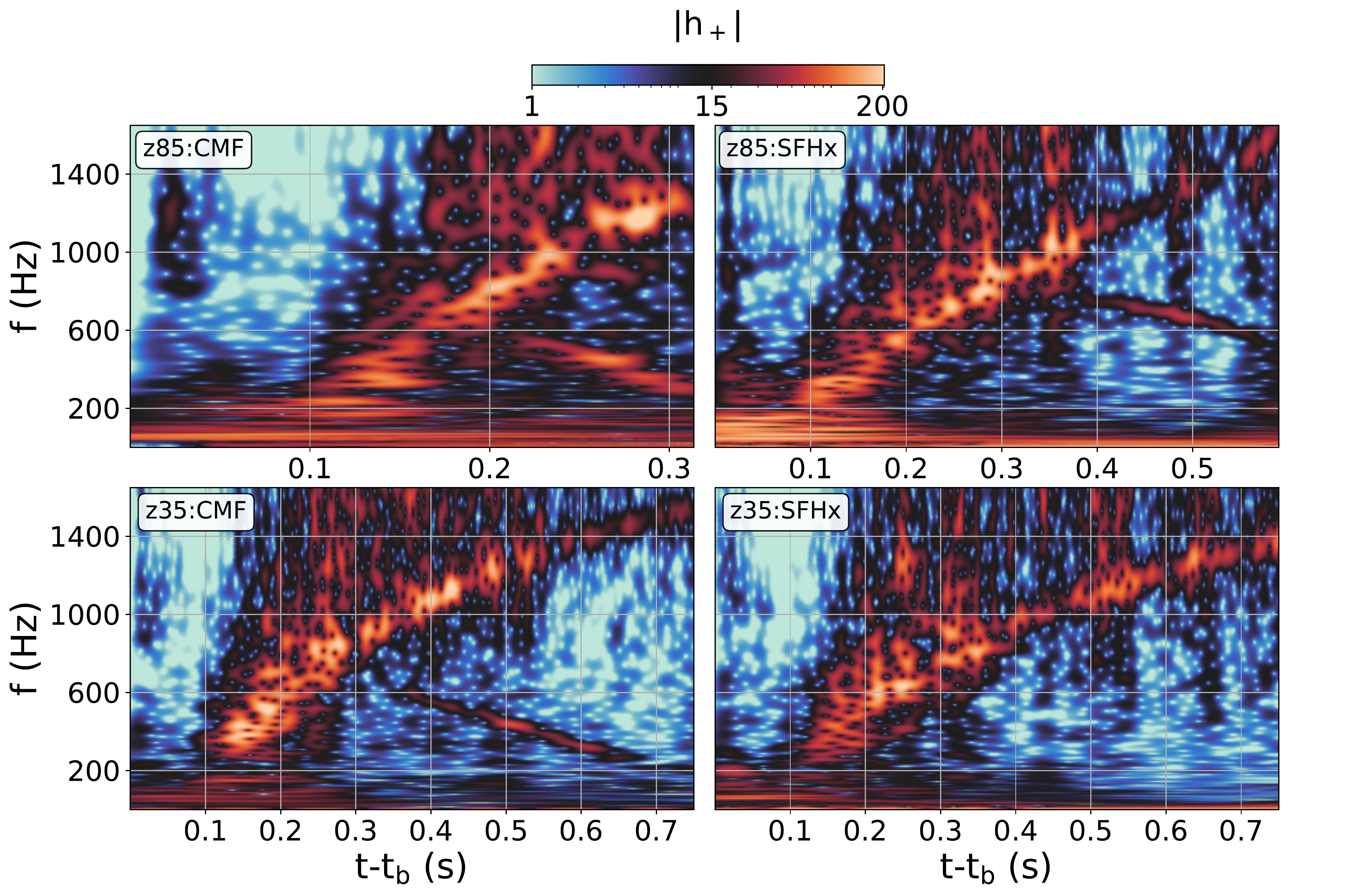}
    \caption{
    GW  spectrograms for 
    \texttt{z85} (top) and \texttt{z35} (bottom) using the \texttt{CMF}-EoS (left) and \texttt{SFHx}-EoS (right). 
    The same logarithmic color scale for the amplitude
    $|h^+|$ is used for all models.
          Models \texttt{z85:CMF}, \texttt{z85:SFHx} and \texttt{z35:CMF} exhibit a distinct second frequency band 
          from the ${}^2\!g_1$-mode,
          which branches off the dominant band after a few hundred  milliseconds. }
    \label{fig:gw_all}
\end{figure}

\begin{figure*}
    \centering
    \includegraphics[width=2.07\columnwidth]{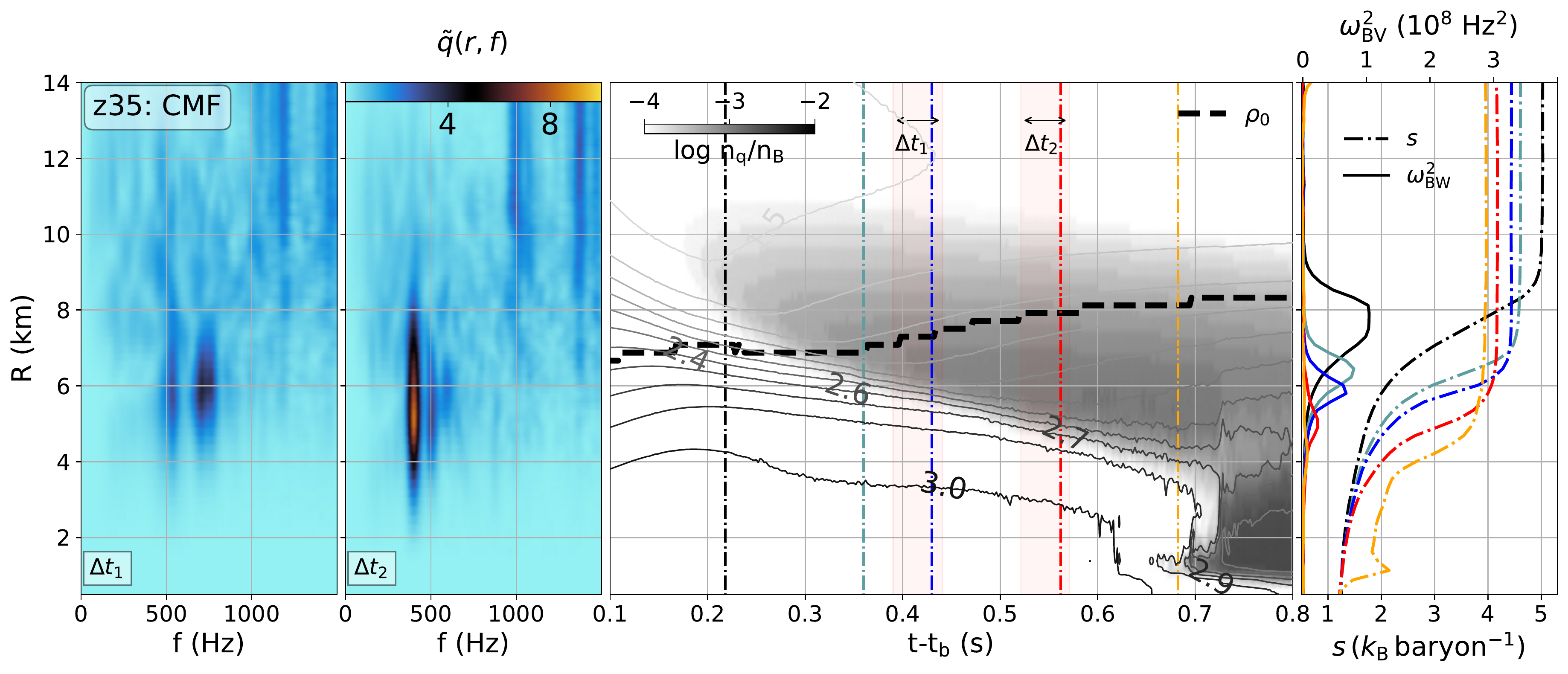}
    \caption{\textit{First and second panel:} 
     Amplitude $\tilde{q}(r,f)$ of the Fourier transform of the quadrupolar perturbation $q(r,t)$ as a function of radius and frequency $f$ for the time intervals $\Delta t_1$ and $\Delta t_2$ around $0.4\, \mathrm{s}$ and $0.55\, \mathrm{s}$ (indicated in red shades in the third panel) for model \texttt{z35:CMF}.
     \textit{Third panel:} Color contour plot of the quark fraction $n_\text{q}/n_\text{B}$  as a function of time and radius, combined with isocontours for the adiabatic index (solid lines in different shades of gray) and a thick black line indicating the radius corresponding to nuclear saturation density.
     \textit{Fourth panel:} \bv $\omegabv$ (solid lines) and spherically averaged specific entropy per baryon
     $s$, (dashed) at five different times, which are indicated on the time axis of the third panel as vertical lines of the respective colour. The last entropy profile (orange) shows a blip at a radius of less than $2\, \mathrm{km}$, which is due to a convective plume that penetrates the PNS core when the buoyancy barrier at the inner edge of the PNS convection zone is eroded ($\omegabv\approx 0$).
       }
    \label{fig:julia_z35_cmf}
\end{figure*}

\textit{Results.}---
Dynamically, the \texttt{CMF} and \texttt{SFHx} models exhibit similar behaviour. Both \texttt{z85} models undergo shock revival followed by early black hole formation, albeit earlier by more than $0.2\, \mathrm{s}$ in \texttt{z85:CMF}. Both \texttt{z35} models 
explode. The GW signals of the \texttt{CMF} and \texttt{SFHx} models exhibit distinctive differences, however.
Figure~\ref{fig:gw_all} shows GW spectrograms computed using
the
\href{https://docs.scipy.org/doc/scipy/reference/generated/scipy.signal.morlet.html}{Morlet wavelet transform} \citep{Morlet_1982}.

The early phase of GW emission is still similar for both EoSs. The \texttt{z85} models show low-frequency emission 
at $\mathord{\sim} 100\,\text{Hz}$ 
due to prompt convection and early SASI activity
\citep{Marek_2009,Murphy_2009,Yakunin_2010,mueller_2013}; this is largely absent in the \texttt{z35} models. Subsequently, the PNS surface $f/g$-mode
\citep{mueller_2013,Morozova_2018,Torres_2018,torres_2019} appears as a prominent emission band with frequencies that increase from $\mathord{\sim} 300\,\text{Hz}$ to above $1000\,\text{Hz}$. The $f/g$-mode frequency rises slightly faster in the \texttt{CMF} models.

The most striking differences are found in another emission band of \emph{decreasing} frequency that branches off the dominant $f/g$-mode between
$0.2\,\mathrm{s}$ and $0.35\, \mathrm{s}$, except in \texttt{z35:SFHx} which shows no such signal. A linear mode analysis (see \cite{Morozova_2018,torres_2019,Andersen_2021,sotani_21} for the methodology) identifies this frequency band as the decreasing branch of the
${}^2\!g_1$ mode (Zha et al.\ in prep.), i.e., a quadrupolar $g$-mode with one node, with an eigenfunction mostly confined to the PNS core (core $g$-mode). Henceforth we refer to the decreasing branch 
as the ${}^2\!g_1$ mode for short\footnote{The dominant band with increasing frequency follows the \emph{increasing} branch of the ${}^2\!g_1$ mode initially and
then the $f$-mode after the avoided crossing of the two modes. The mode classification is, e.g., sensitive to the boundary condition in the linear analysis.}.

The mode frequency $f_{{}^2\!g_1}$ is systematically lower in \texttt{z85:CMF}  compared to \texttt{z85:SFHx}. 
In \texttt{z85:CMF}, $f_{{}^2\!g_1}$ decreases from $\mathord{\sim} 600 \,\mathrm{Hz}$ at $0.2\,\mathrm{s}$ to $\mathord{\sim} 220\,\mathrm{Hz}$ at $0.32\,\text{s}$, at which point the model collapses to a BH. In \texttt{z85:SFHx}, BH collapse occurs later and $f_{{}^2\!g_1}$  evolves more slowly from a higher frequency of $\mathord{\sim}800\,\mathrm{Hz}$ down to $\mathord{\sim} 560\,\mathrm{Hz}$ at $0.58\,\text{s}$. In \texttt{z35:CMF}, the 
${}^2\!g_1$ mode lives at similarly low frequencies as in \texttt{z85:CMF}, i.e., in the range $220\texttt{-}600\,\mathrm{Hz}$.

Such pronounced emission in the declining ${}^2\!g_1$-mode frequency band as in the \texttt{CMF} models (and to a lesser extent model \texttt{z85:SFHx}) is not usually observed in simulations with energy-dependent neutrino transport.
These usually show an emission \emph{gap} at the avoided crossing with the $f$-mode \cite{Morozova_2018}.
The ${}^2\!g_1$-mode has been found in simulations with more approximate neutrino transport \cite{pablo_2013,Kawahara_2018}, or modified Newtonian gravity \cite{torres_2019,torres_2019b,vartanyan_23}. 

To further confirm the nature of the mode,
we perform a spatially resolved Fourier analysis
of the integrand of the quadrupole formula using high-time-resolution simulation output with sampling frequency $10^4\,\text{Hz}$. To detect quadrupolar motions as a function of radius and frequency, we integrate over angle only, and obtain a radius-dependent measure $q(r,t)$ of quadrupolar perturbations,
\begin{align}\label{eq:qa}
    q(r,t) & = \, \frac{32\pi^{3/2}G}{\sqrt{15}\,c^4} \int_0^{\pi} \,\D\theta\, \phi^6 r^3\sin\theta\nonumber \\
    & \times \left\{ \,\pdv{}{t} \left(S_{r}(3\cos^2\theta -1)\right) + \frac{3}{r} S_\theta\sin\theta\cos\theta\right\}.
\end{align}
$\phi$ is the conformal factor of the space-time metric, 
and $S_r$ and $S_\theta$ are the orthonormal components of the relativistic momentum density.

We obtain spectrograms of $q(r,t)$ (Figure~\ref{fig:julia_z35_cmf}, first two panels)  using the \href{https://juliamath.github.io/FFTW.jl/stable/}{Fast-Fourier transforms} (FFT) in a fixed time window $\Delta t$ and apply additional denoising by convolving the FFT with a weighted sum of \href{https://github.com/francescoalemanno/KissSmoothing.jl}{radial basis functions} \citep{myers_1970}.

Spectrograms are shown for two time windows $\Delta t_1$ and $\Delta t_2$ around
$0.4\,\mathrm{s}$ and $0.55\,\mathrm{s}$ (marked as red shaded areas in the third panel of Figure~\ref{fig:julia_z35_cmf}) for model \texttt{z35:CMF}.
During $\Delta t_1$, the specrograms show power corresponding to the low-frequency signal at $\approxinbetween{r}{4}{8}{km}$  and a frequency of $\mathord{\sim} 600\,\mathrm{Hz}$, with a weaker ``hotspot'' at $\mathord{\sim} 500\,\mathrm{Hz}$. Later, during $\Delta t_2$, the hotspot is stronger and its centroid shifts towards small radii (although it still reaches out to $\mathord{\sim}\,8\,\mathrm{km}$) as the PNS contracts. The frequency decreases to $\mathord{\sim} 430 \,\mathrm{Hz}$ and
is clearly defined.  Profiles of the relativistic Brunt-V\"ais\"al\"a frequency $\omegabv$ \citep{mueller_2013} and specific entropy per baryon $s$
(Figure~\ref{fig:julia_z35_cmf}, fourth panel) show that the mode is located at the \emph{inner} boundary of the PNS convection zone, originating from a different region than the high-frequency emission, which is visible in two streaks above $1\,\mathrm{kHz}$ at larger radii.  The profiles of $\omegabv$ also explain the downward trend in frequency as the peak in 
$\omegabv$ decreases with time. All of this strongly supports the identification as a core $g$-mode.

This still leaves the question why the  
${}^2\!g_1$-mode has a significantly lower frequency and is more strongly excited in the \texttt{CMF} models. Before a more quantitative analysis, it is important to note that quark formation is not \emph{directly} responsible  in the \texttt{CMF} models for the smaller mode frequency as evident from the time- and radius-dependent quark fraction $n_\mathrm{q}/n_\mathrm{B}$ (Figure~\ref{fig:julia_z35_cmf}, third panel).

Although quarks appear off-center at $\approxinbetween{r}{8}{10}{km}$ quite early at $\mathord{\sim} 0.2\,\mathrm{s}$, they appear only in small numbers $n_\mathrm{q}/n_\mathrm{B}\leq 10^{-4}$.
The appearance of quarks at low densities
is due to the absence of a sharp phase transition in the \texttt{CMF} EoS and the high temperatures in the PNS mantle, but is of little dynamical relevance. Quarks only appear more abundantly and lead to significant softening later at $\mathord{\sim} 0.7\,\mathrm{s}$ at radii $\mathord{\sim} 2\,\mathrm{km}$. Thus, the full transition to quark matter comes too late to account for the distinct ${}^2\!g_1$-mode  in \texttt{z35:CMF}
as opposed to \texttt{z35:SFHx} well before $0.7\,\mathrm{s}$.

\begin{figure}
	\includegraphics[width=\columnwidth]{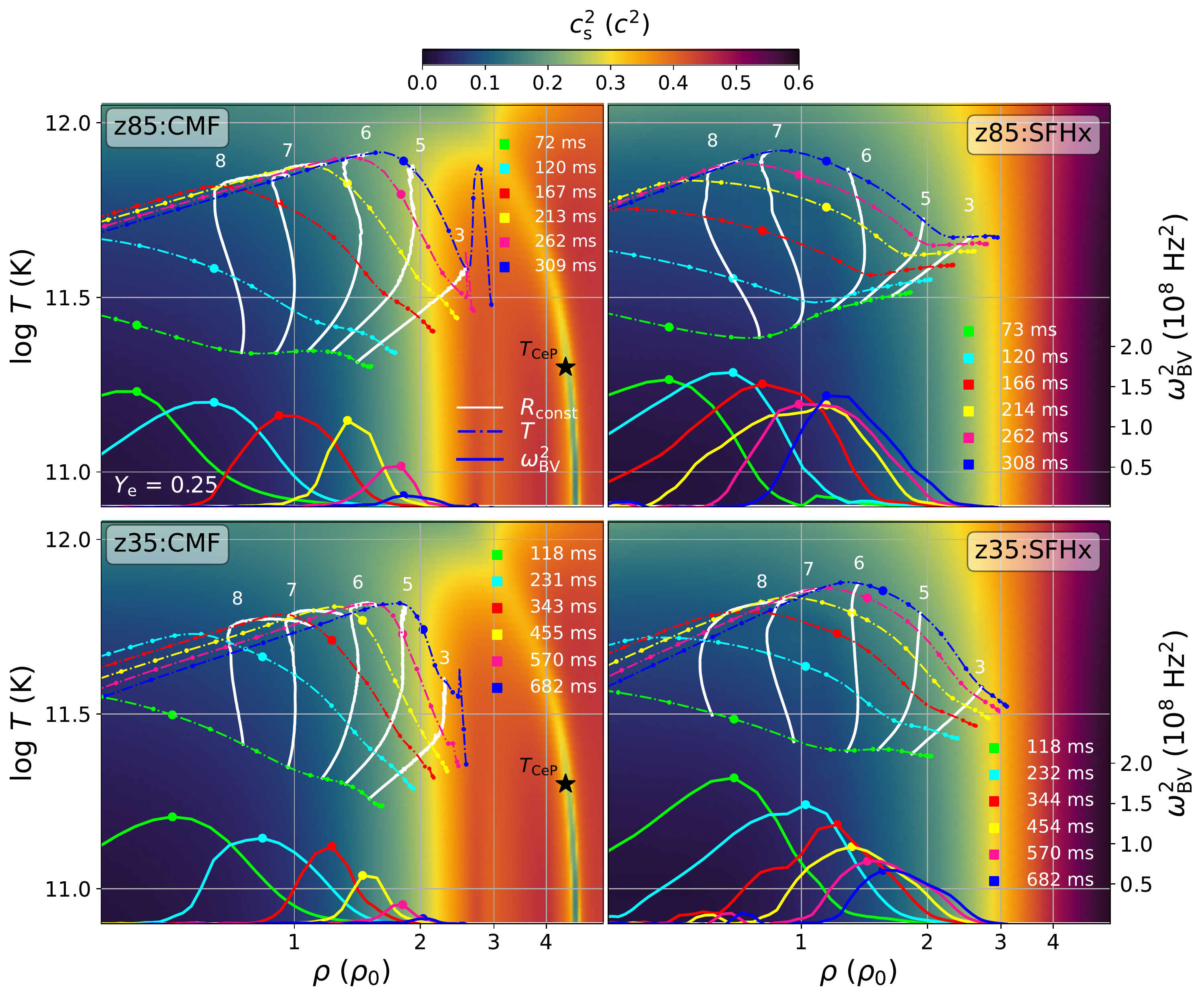}
    \caption{Color contour plots of the squared sound speed $c_\mathrm{s}^2$ as a function of density $\rho$ and temperature $T$ for an electron fraction of $Y_\text{e}=0.25$ for models $\texttt{z85}$ (top) and $\texttt{z35}$ (bottom) with the \texttt{CMF} EoS (left), and the \texttt{SFHx} EoS (right). Solid and dashed curves show $\omegabv$ and (spherically averaged) temperature profiles $T(\rho)$    
    at various times (indicated by line color). The density where $\omegabv$ peaks is indicated by a dot on the curves.
    The white lines indicate five constant radii; the radius (in km) is indicated on top. 
    Note a temperature blip at high densities due to the overshooting of plumes into the PNS at late times (blue curves) in the \texttt{CMF} models.}
    \label{fig:speefOfSound_both}
\end{figure}

The lower ${}^2\!g_1$-mode frequency in the 
 \texttt{CMF} models is rather connected to
lower peaks of $\omegabv$ at the inner boundary of the PNS convection zone at densities below $2\mathord{\times}\rho_0$ (colored solid lines in Figure~\ref{fig:speefOfSound_both}). At late times the buoyancy barrier at the bottom of the PNS convective zone disappears almost entirely, and the entropy profiles show overshooting into the core as favourable conditions for ``inverted convection''  to develop~\citep{jakobus_2022}.
The reason for the lower \bv can be analyzed by writing $\omega_\mathrm{BV}^2$ as
\begin{align}\label{eq:bv_rewr}
    \omegabv{} = \dv{\alpha}{r}\frac{\alpha}{\rho h \phi^4} \frac{1}{c_\mathrm{s}^2}\left[\left(\pdv{ P}{s}\right)_{\!\!\tilde{\rho},Y_e}\dv{s}{r}+\left(\pdv{P}{Y_e}\right)_{\!\tilde{\rho},s}\dv{Y_e}{r}\right], 
\end{align}
and considering the impact of the various gradients and thermodynamic derivatives.
Here, $\alpha$ is the lapse function, $\rho$ is the baryonic mass density, $\tilde{\rho}$ is the total mass-energy density, $P$ is the pressure, $h=(\tilde{\rho}+P/c^2)/\rho$ is the relativistic enthalpy, $Y_\mathrm{e}$ is the electron fraction, and $c_\mathrm{s}$ is the sound speed.
The most conspicuous feature in the \texttt{CMF} models
is a higher sound speed than in the \texttt{SFHx} models at the location corresponding to the maximum of  $\omega_\mathrm{BV}$ (color shading in Figure~\ref{fig:speefOfSound_both}), once the inner edge of the PNS convection zone contracts to densities of about $2\mathord{\times}\rho_0$. This is due to significant stiffening
of the \texttt{CMF} EoS owing to baryon-baryon repulsion~\citep{motornenko_2020}. Although not directly related to the formation of quark matter, such pronounced stiffening at moderately high densities is characteristic of currently viable EoS with a phase transition or crossover to quark matter \cite{altiparmak_22,Fujimoto:2019hxv, Soma:2022vbb}. The stiffening is crucial for achieving maximum neutron star masses compatible with observational constraints and tentatively supported by heavy-ion collisions  \cite{kuttan_22,Oliinychenko:2022uvy}. On top of the systematic difference in sound speed between the two EoSs, we also find a somewhat disparate PNS structure, which complicates the comparison of $\omegabv$ between  
\texttt{CMF} and \texttt{SFHx} models, e.g., the inner edge of the PNS convection zone as defined by the peak in $\omega_\mathrm{BV}$ tends to lie at higher densities in the \texttt{CMF} models. 
Although the difference in sound speed has a clear impact on $\omega_\mathrm{BV}$ and can be readily connected to the underlying physics of the EoS, there are further smaller effects that will eventually need to be incorporated in a rigorous theory for the EoS-dependence of the ${}^2\!g_1$-mode.
Differences in the electron fraction gradient $\mathrm{d}Y_\text{e}/\mathrm{d}r$ also contribute to the lower $\omega_\mathrm{BV}$ for \texttt{CMF}. Different from
the \texttt{SFHx} models, $\mathrm{d}Y_\text{e}/\mathrm{d}r$ 
becomes negative in the region of interest before the onset of the signal. With positive $\left(\partial P/\partial Y_\mathrm{e}\right)_{\tilde{\rho},s}$, the  second term in brackets in Equation~(\ref{eq:bv_rewr}) then decreases $\omega_\mathrm{BV}$, especially since $\left(\partial P/\partial Y_\mathrm{e}\right)_{\tilde{\rho},s}$ diverges from the SFHx EoS at this point and becomes larger by up to a factor of four in the PNS core in the \texttt{CMF} models. The ultimate cause for the different behaviour is that the small ``hump'' in $Y_\mathrm{e}$ at a mass coordinate of $\mathord{\sim}0.7\,M_\odot$  that is imprinted on the PNS structure shortly after bounce is erased quicker by neutrino diffusion in the \texttt{CMF} models. The terms $\left(\partial P/\partial s\right)_{Y_\mathrm{e},\tilde{\rho}}$ and $\mathrm{d}s/\text{d}r$ also show some EoS dependence, but their net effect is even smaller.
Details are shown in the Supplementary Material.

It is more challenging to trace the higher \emph{power} in GWs emitted by the ${}^2\!g_1$-mode to the PNS structure and to EoS properties. Stronger excitation of the ${}^2\!g_1$-mode in the \texttt{CMF} models could be due to stronger PNS convection or more efficient coupling between the forcing convective motions and the  ${}^2\!g_1$-mode. Stronger convection in the \texttt{CMF} models appears to at least play an important role. The turbulent kinetic energy in the PNS convection zone is about an order of magnitude larger in \texttt{z35:CMF} at several $10^{50}\,\mathrm{erg}$ than in \texttt{z35:SFHx},
and still somewhat larger in \texttt{z85:CMF} than in \texttt{z85:SFHx} (see Supplementary Material). Both \texttt{z85} models have significantly higher turbulent convective energies than the \texttt{z35} models in line with recent findings of stronger PNS convection for more massive progenitors \cite{nagakura_2020}.
The higher convective energies are mostly due to higher turbulent velocities and less due to differences in the mass of the PNS convection zone. The empirical finding of stronger PNS convection in the \texttt{CMF} models explains the conspicuous signal from the ${}^2\!g_1$-mode in the GW spectrograms, but further work is needed to identify the underlying physical reason. Unfortunately, the dynamics of PNS convection are not fully understood because of the complicated interplay of stabilising and destabilising stratification gradients and multi-dimensional convective flow \cite[e.g.,][]{bruenn_96,bruenn_04,glas_2019,powell_2019,mueller_2020}. It is noteworthy, however, that a significant impact of EoS properties on the long-term behaviour of PNS convection during the Kelvin-Helmholtz cooling phase has been reported before \cite{roberts_12}.

\textit{Conclusions.}---
Our 2D supernova simulations with a quark-hadron \texttt{CMF} EoS
\cite{motornenko_2020} and the hadronic \texttt{SFHx} EoS \cite{Steiner_2013} show a characteristic GW emission band with \emph{decreasing} frequencies of several hundred Hz in addition to the well-known emission band
from the dominant $f$/$g$-mode. We identified a core $g$-mode (${}^2\!g_1$-mode)
that mostly lives around the inner boundary of the PNS convection zone as the oscillation mode responsible for this GW feature.

The mode frequency and power are very sensitive to the high-density EoS. For a $35\,\mathrm{M}_\odot$
progenitor, the GW signal from the ${}^2\!g_1$-mode is only present for the \texttt{CMF} EoS, and for a $85\,\mathrm{M}_\odot$ progenitor, it is stronger, appears earlier and lies at lower frequencies for the \texttt{CMF} EoS. The lower frequency indicates a softening of the inner boundary of the PNS convection zone primarily due to a higher sound speed of the \texttt{CMF} EoS at densities of about $2\mathord{\times}\rho_0$.
The strength of PNS convection as a driver of the ${}^2\!g_1$-mode
is sensitive to the EoS, which explains the stronger GW signal from this mode for the \texttt{CMF} EoS.

These results suggest that the supernova GW signal holds more promise for probing properties of nuclear matter beyond saturation density than hitherto thought because the signal from the ${}^2\!g_1$-mode is determined by the behaviour of the EoS around $2\mathord{\times}\rho_0$, in contrast to the dominant $f/g$-mode, which lives primarily at the PNS surface and is determined by bulk PNS parameters. The use of the ${}^2\!g_1$ mode as a probe for the high-density EoS is not limited to the scenario of
a first-order phase transition considered in earlier work \cite{McDermott_1990,Strohmayer_1993,Miniutti_2003}.
The signal from the ${}^2\!g_1$-mode still cannot probe quark formation directly, but may be used to measure the stiffness of the EoS in the aforementioned density regime, which will have implications for the viability of a phase transition or smooth crossover to quark matter. Observations of the  ${}^2\!g_1$-mode feature from a Galactic supernova could thus complement heavy-ion collisions \citep{huth_2022} and current astrophysical constraints on the stiffness of the high-density EoS
from pulsar masses \citep{Demorest_2010}
and NICER data \cite{riley_2021,Miller_2021} and the GW signal from neutron star mergers. The tidal deformability parameter from the pre-merger GW signal from GW170817 already rules out very stiff EoS~\citep{Abbott_2017,abbott_2018}. In contrast, recent radius measurements by NICER argue against substantial softening of matter between $2$--$3\mathord{\times} \rho_0$ and $4$--$5\mathord{\times}\rho_0$ (that would accompany a strong first-order PT in this density regime)~\citep{Kojo_2022}. 
Future work should explore the impact of the EoS and of dimensionality on mode excitation and the signal from the ${}^2\!g_1$-mode more broadly. Encouragingly, despite generally lower GW amplitudes in three dimensions (3D), the signal has been found in 3D models after submission of our paper \cite{vartanyan_23}. One should also further clarify the physical parameters that govern the mode frequency and power and assess the potential of current and next-generation GW interferometers to detect the signal and reconstruct the trajectory of the mode frequency.

 \textit{Acknowledgements}---
The authors are supported by the Australian Research Council (ARC) Centre of Excellence (CoE) for Gravitational Wave Discovery (OzGrav) project number CE170100004.  BM is supported by ARC Future Fellowship FT160100035.  
AH is supported by the ARC CoE for All Sky Astrophysics in 3 Dimensions (ASTRO 3D), through project number CE170100013.
JP is supported by the ARC Discovery Early Career Researcher Award (DECRA) project number DE210101050. 
AM acknowledges the Stern–Gerlach Postdoctoral fellowship and the educator program of the Stiftung Polytechnische Gesellschaft.
HS thanks the Walter Greiner Gesellschaft zur F\"orderung der
physikalischen Grundlagenforschung e.V. through the Judah M. Eisenberg
Laureatus Chair at the Goethe Universit\"at Frankfurt am Main. 
We acknowledge computer time allocations from Astronomy Australia Limited's ASTAC scheme, the National Computational Merit Allocation Scheme (NCMAS), and
from an Australasian Leadership Computing Grant.
Some of this work was performed on the Gadi supercomputer with the assistance of resources and services from the National Computational Infrastructure (NCI), which is supported by the Australian Government.  Some of this work was performed on the OzSTAR national facility at Swinburne University of Technology.  OzSTAR is funded by Swinburne University of Technology and the National Collaborative Research Infrastructure Strategy (NCRIS). Some of
the computational resources were provided by the Goethe-HLR computing
center.

\bibliography{cit}


\end{document}


\title{Supplementary Material:\\ Gravitational Waves from a Core g-Mode in Supernovae as Probes of the High-Density Equation of State}
\date{\today}
\author{Pia Jakobus}
\email{pia.jakobus@monash.edu}
\author{Bernhard M\"uller}
\author{Alexander Heger}
\affiliation{School of Physics and Astronomy, Monash University, Clayton, VIC~3800 Australia}
\author{Shuai Zha}
\affiliation{Tsung-Dao Lee Institute, Shanghai Jiao Tong University, Shanghai 200240, China}
\author{Jade Powell}
\affiliation{Centre for Astrophysics and Supercomputing, Swinburne University of Technology, Hawthorn, VIC 3122, Australia}
\author{Anton Motornenko}
\author{Jan Steinheimer}
\author{Horst Stoecker}
\affiliation{Frankfurt Institute for Advanced Studies, Giersch Science Center, Frankfurt am Main, Germany}
\maketitle

\section{Turbulent kinetic energy in the convection zone}
We define the turbulent kinetic energy
in terms of the turbulent velocity fluctuations
$\delta v$ and the density $\rho $ as
\begin{equation}
E_\text{kin}^\text{turb} = \frac{1}{2}\int_\text{PCS} \delta v^2 \rho \mathrm{d}V,
\end{equation}
where the volume element $\mathrm{d}V$ includes the appropriate relativistic metric factors\footnote{The deviation of the Lorentz factor from unity is negligible in the proto-compact star because the fluid velocities are small.}, i.e., $\mathrm{d} V=2\pi \phi^6 r^2 \sin\theta \,\mathrm{d}r\, \mathrm{d}\theta$ in axisymmetry with the conformal factor $\phi$. The integral is performed over the entire proto-compact star ($\mathrm{PCS}$), i.e., for densities above $10^{11}\, \mathrm{g}\,\mathrm{cm}^{-3}$. The turbulent velocity fluctuations are given by
\begin{equation}
\delta v^2=(v_r-\langle v_r\rangle)^2+
v_\theta^2,
\end{equation}
where $\langle v_r\rangle$  is the spherical Favre average of the radial velocity.

\begin{figure}%
    \centering
    \subfloat[\centering ]{{\includegraphics[width=8cm]{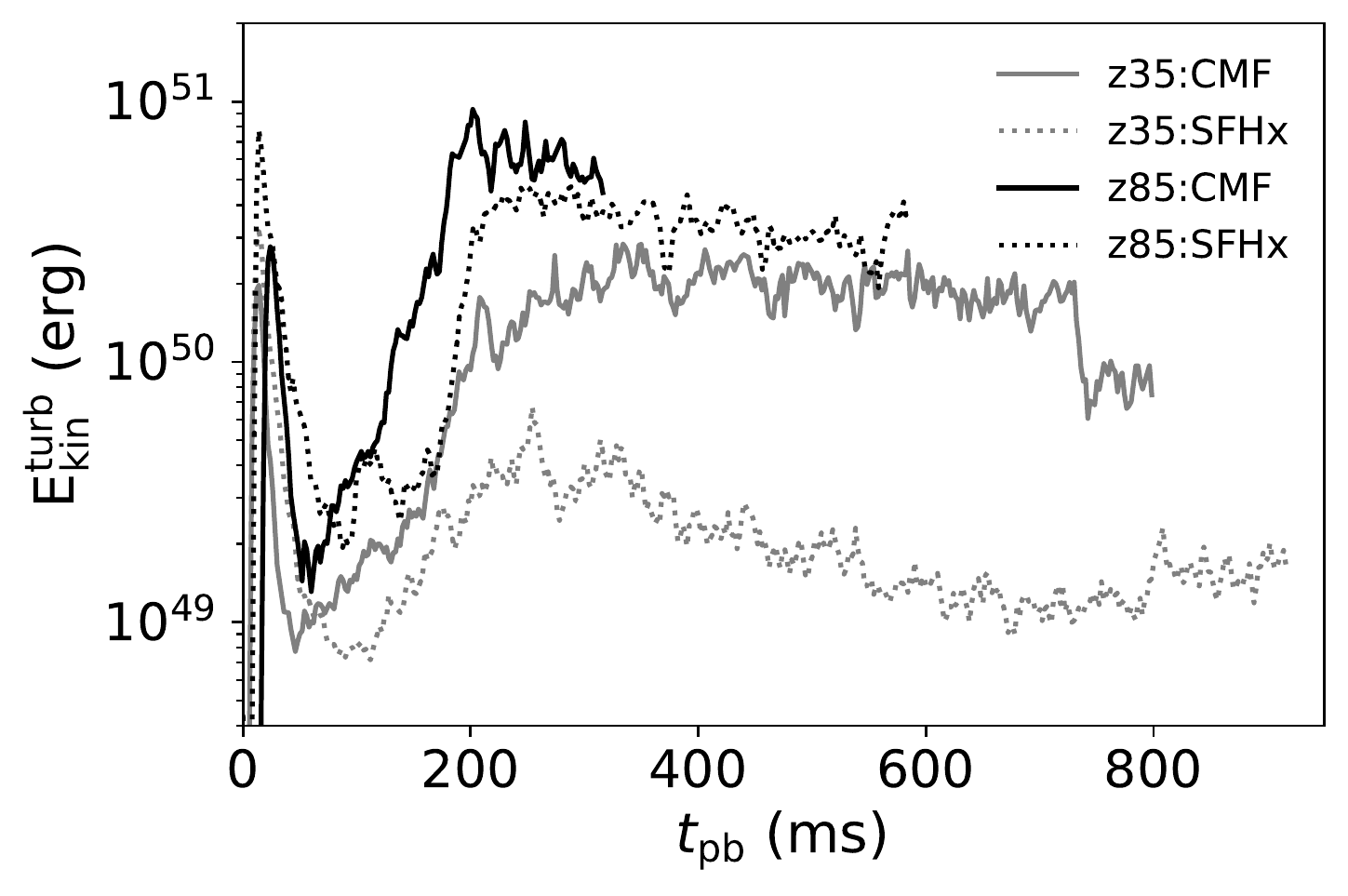} }}%
    \qquad
    \subfloat[\centering ]{{\includegraphics[width=8cm]{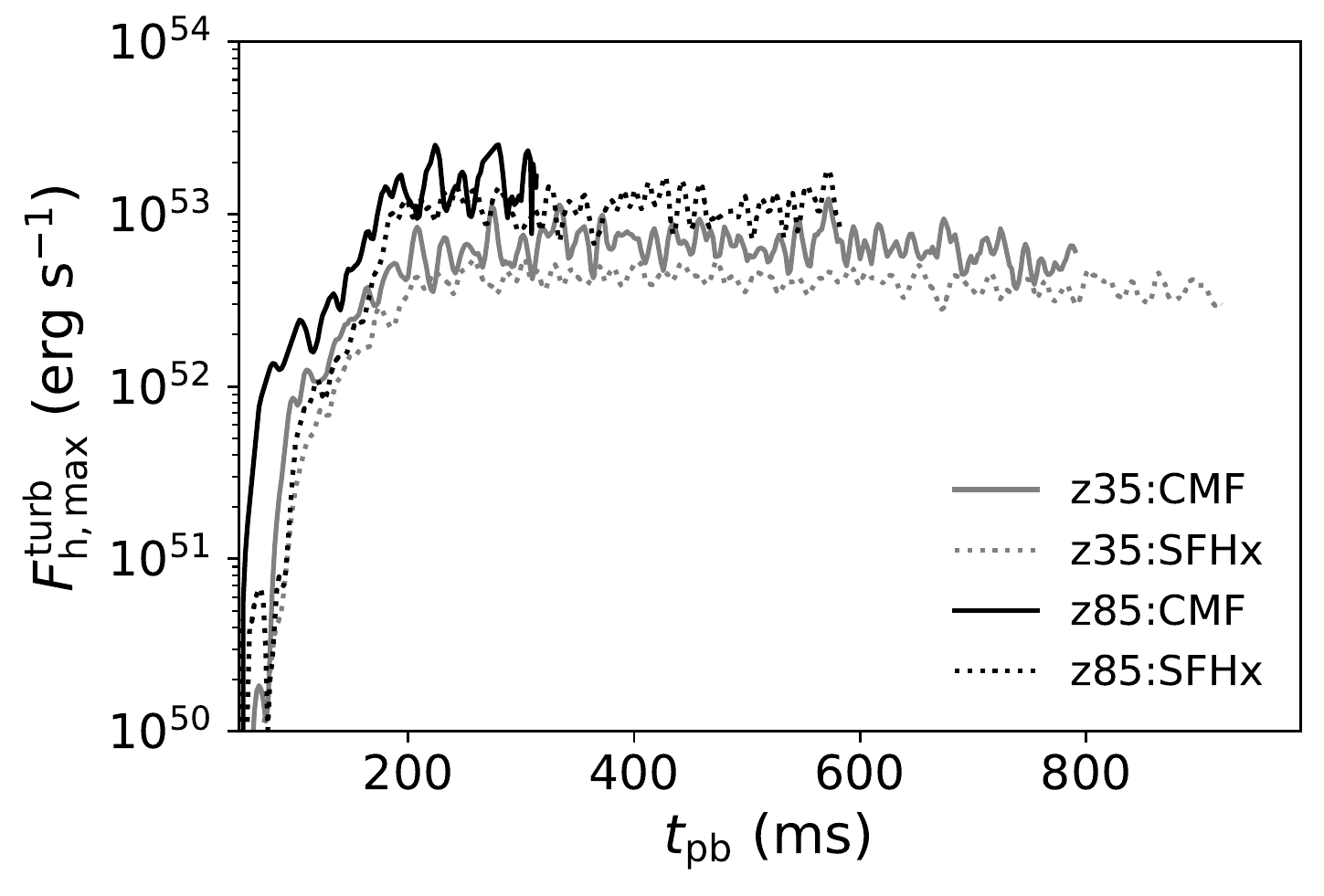} }}%
    \caption{(a) Turbulent kinetic energy of the PCS as function of post-bounce time for the progenitors \texttt{z35} (grey) and \texttt{z85} (black) using the \texttt{SFHx} EoS (dashed lines) and \texttt{CMF} EoS (solid lines). (b) Maximal turbulent energy flux $F_\mathrm{conv,max}^\mathrm{h}$ in the PCS convection zone (time averaged over 10 ms) for all four models as a function of post-bounce time. The same line styles as in Panel~(a) are used.}%
    \label{fig:ekin turb}%
\end{figure}

We plot $E_\text{kin}^\text{turb}$ for all four models as function of time in Figure~\ref{fig:ekin turb}. The \texttt{CMF} models show higher turbulent kinetic energies compared to the \texttt{SFHx} models, especially for the lighter \texttt{z35} progenitor. In this case, the turbulent kinetic energy in the PCS convection zone is about an order of magnitude higher in the \texttt{CMF} EoS compared to \texttt{SFHx}. For \texttt{z85}, the difference is modest but robust, and the rise of the turbulent kinetic energy to the plateau at $\sim 5\times 10^{50}\, \mathrm{erg}$ occurs significantly earlier.\\
In addition, we consider the turbulent energy flux as
\begin{align}
    {F}_\mathrm{turb}^\mathrm{h} &=  4\pi r^2 \langle\rho {v}_r'' {h}'' \rangle\\
&= 4\pi r^2 \int \alpha\phi^4 \rho{v}_r''{h}''r^2\mathrm{d}\Omega,
\end{align}
where $X''=X-\langle X \rangle$, with the Favre average $\langle X \rangle$. $\alpha$ is the lapse function. The behaviour of the turbulent flux is similar to the turbulent kinetic energy: \texttt{z85:CMF} exhibits the largest flux, followed by \texttt{z85:SFHx}, and \texttt{z35:CMF}. 
The progenitor \texttt{z35:SFHx} exhibits a notably lower flux, although the difference in comparison to \texttt{z35:CMF} is relatively smaller than what is observed for the turbulent kinetic energy in the left plot of Fig.~\ref{fig:ekin turb}. High peak values are reached by this metric for
the strength of convection i) because  the neutrino losses that drive convection are unusually for these high mass progenitors, and ii) because this metric is sensitive to considerable fluctuations in the profiles of the turbulent energy flux in 2D simulations.

\section{Electron fraction gradient}
\begin{figure}[H]
\centering 
	\includegraphics[width=\columnwidth]{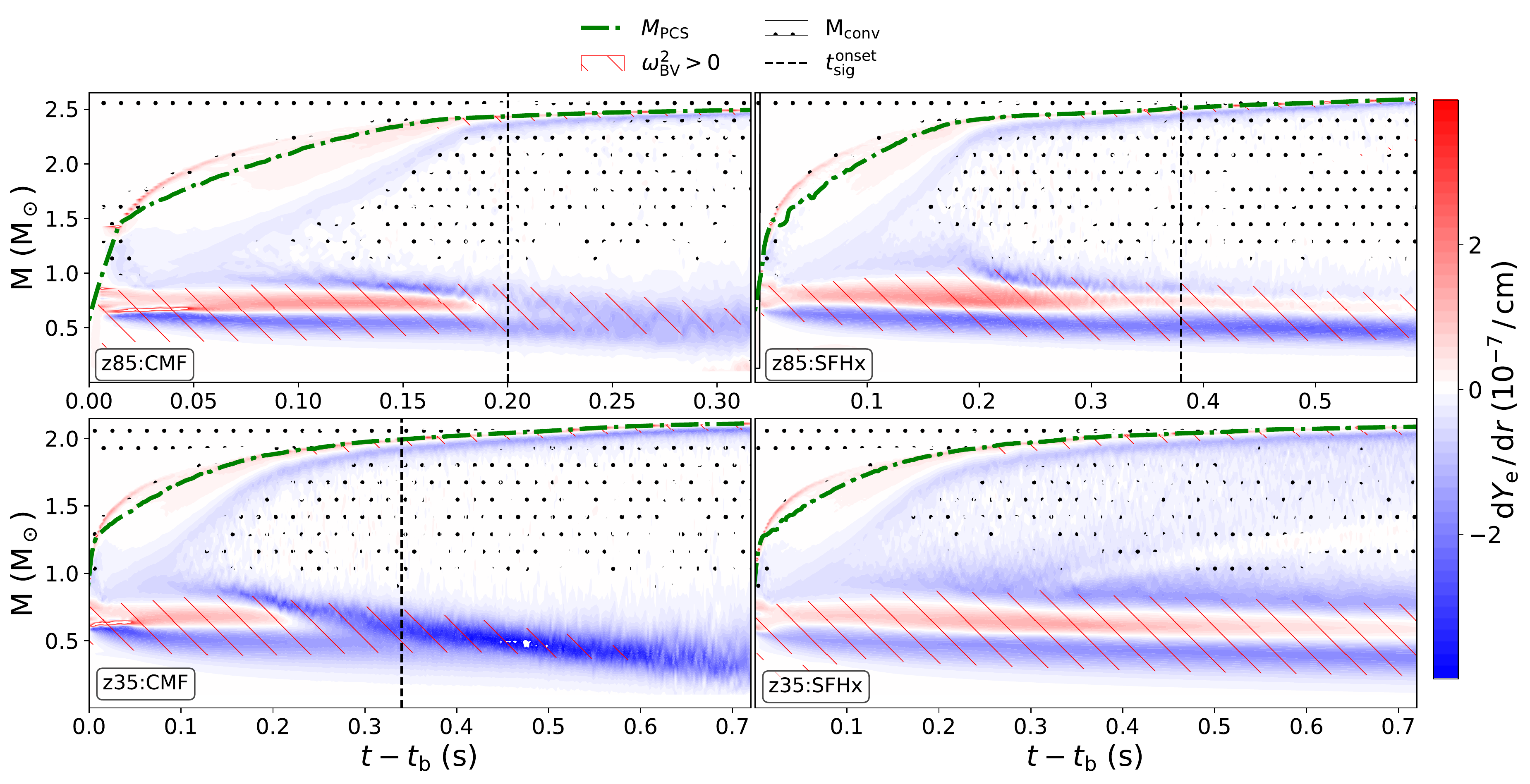}
    \caption{Evolution of the gradient $\mathrm{d}Y_e/\mathrm{d}r$
    of the spherically-averaged electron fraction
    as function of post-bounce time for the \texttt{CMF} model (left), \texttt{SFHx} (right) for \texttt{z85} (upper panel) and \texttt{z35} (lower panel). The black circles approximately track the PCS convection zone. Red dashes show buoyantly stable regions where $\omega_\mathrm{BV}^2 > 0$. Those areas are susceptible to quadrupolar perturbations at the frequency of the $g^2_1$-mode. The black-dashed vertical line roughly corresponds to the onset of the $g^2_1$-signal in the GW spectrograms.}
    \label{fig:dydr}
\end{figure}
The differences in the Brunt-V\"ais\"al\"a frequency in the region below the PCS convection zone in the various models affect the frequency of the $g^2_1$-mode and may also have some bearing on its excitation by turbulent motions in the overlying convection zone. Aside from the different sound speed above nuclear saturation density, differences in the Brunt-V\"ais\"al\"a frequency between the \texttt{CMF} and \texttt{SFHx} models also arise because the electron fraction evolves differently during the post-bounce phase. To illustrate this effect we show
the gradient of the spherically-averaged electron fraction.

Initially, all models show a slight negative electron fraction gradient in the PCS core inside $\mathord{\sim}0.7\,\mathrm{M}_\odot$ (inner blue region in
Figure~\ref{fig:dydr}), and then a small ``hump'' that is visible as a red stripe and blue stripe further outside in Figure~\ref{fig:dydr}. With time, neutrino diffusion erases the hump, and the two blue regions merge in the \texttt{CMF} models. This process takes considerably longer in model \texttt{z85:SFHx}, and in model \texttt{z85:SFHx} the hump is still clearly present at the end of the simulation. The evolution of the electron fraction hump in the code is affected by a combination of factors that determine the neutrino opacities in this region, i.e., density, temperature, and neutron, proton, and neutrino chemical potentials. Hence there is no straightforward explanation for the faster disappearance of the hump in the \texttt{CMF} models at this time.
\newpage 
\section{Decomposition of Terms Contributing to the Brunt-V\"ais\"al\"a frequency}
To elucidate how differences in the entropy and electron fraction gradients and various thermodynamic derivatives affect the Brunt-V\"ais\"al\"a frequency at the inner boundary of the PCS convection zone, we rewrite the relativistic
 Brunt-V\"ais\"al\"a frequency as
\begin{align}\label{eq:bv}
    \omega_\text{BV}^2= \dv{\alpha}{r}\frac{\alpha}{\rho h \phi^4}\frac{1}{c_s^2}
    \left[\left(\pdv{ P}{s}\right)_{\!\!\tilde{\rho},Y_e}\dv{s}{r}+\left(\pdv{P}{Y_e}\right)_{\!\tilde{\rho},s}\dv{Y_e}{r}\right].
\end{align}
Here, $\alpha$ is the lapse function, $\phi$ is the conformal factor, $\rho$ is the baryonic mass density, $\tilde{\rho}$ is the total mass-energy density, $P$ is the pressure, $h=(\tilde{\rho}+P/c^2)/\rho$ is the relativistic enthalpy, $s$ is the specific entropy per baryon, $Y_\mathrm{e}$ is the lepton fraction, and $c_\mathrm{s}^2$ is the (adiabatic) sound speed\footnote{Note that the small effect of neutrino pressure and energy density is not included in Equation~(\ref{eq:bv}).}.

\begin{figure}[H]
\centering 
	\includegraphics[width=0.4\columnwidth]{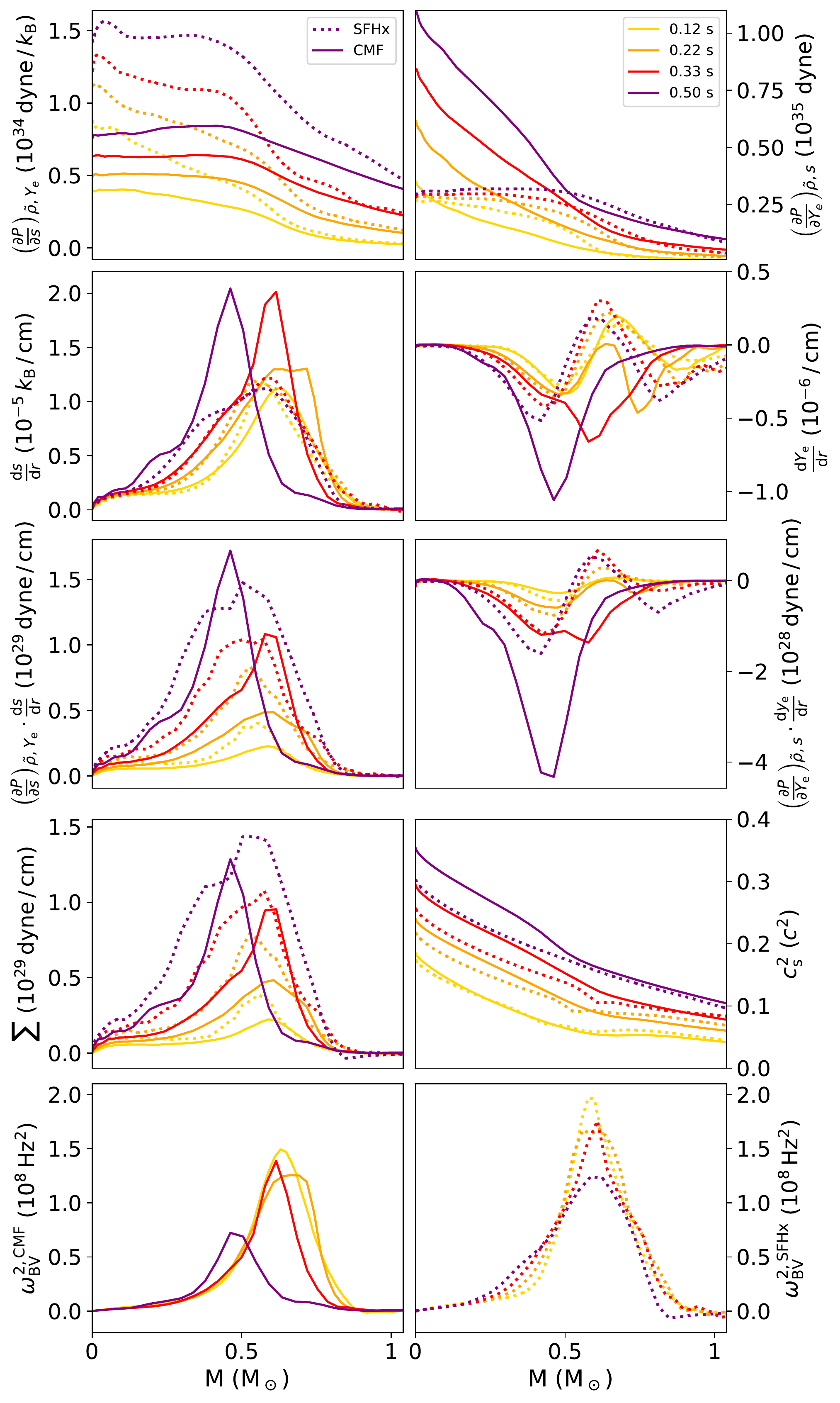}
    \caption{Individual terms in Equation~(\ref{eq:bv}) for \texttt{z35:CMF} (solid lines) and \texttt{z35:SFHx} (dotted) at four different post-bounce times. The left panel in the fourth row shows the term in square brackets of Equation~\ref{eq:bv}, i.g. $\displaystyle{\sum \equiv \left(\pdv{ P}{s}\right)_{\!\!\tilde{\rho},Y_\mathrm{e}}\dv{s}{r}+\left(\pdv{P}{Y_\mathrm{e}}\right)_{\!\tilde{\rho},s}\dv{Y_\mathrm{e}}{r}}$.}
    \label{fig:z35}
\end{figure}

\begin{figure}[H]
\centering 
	\includegraphics[width=0.4\columnwidth]{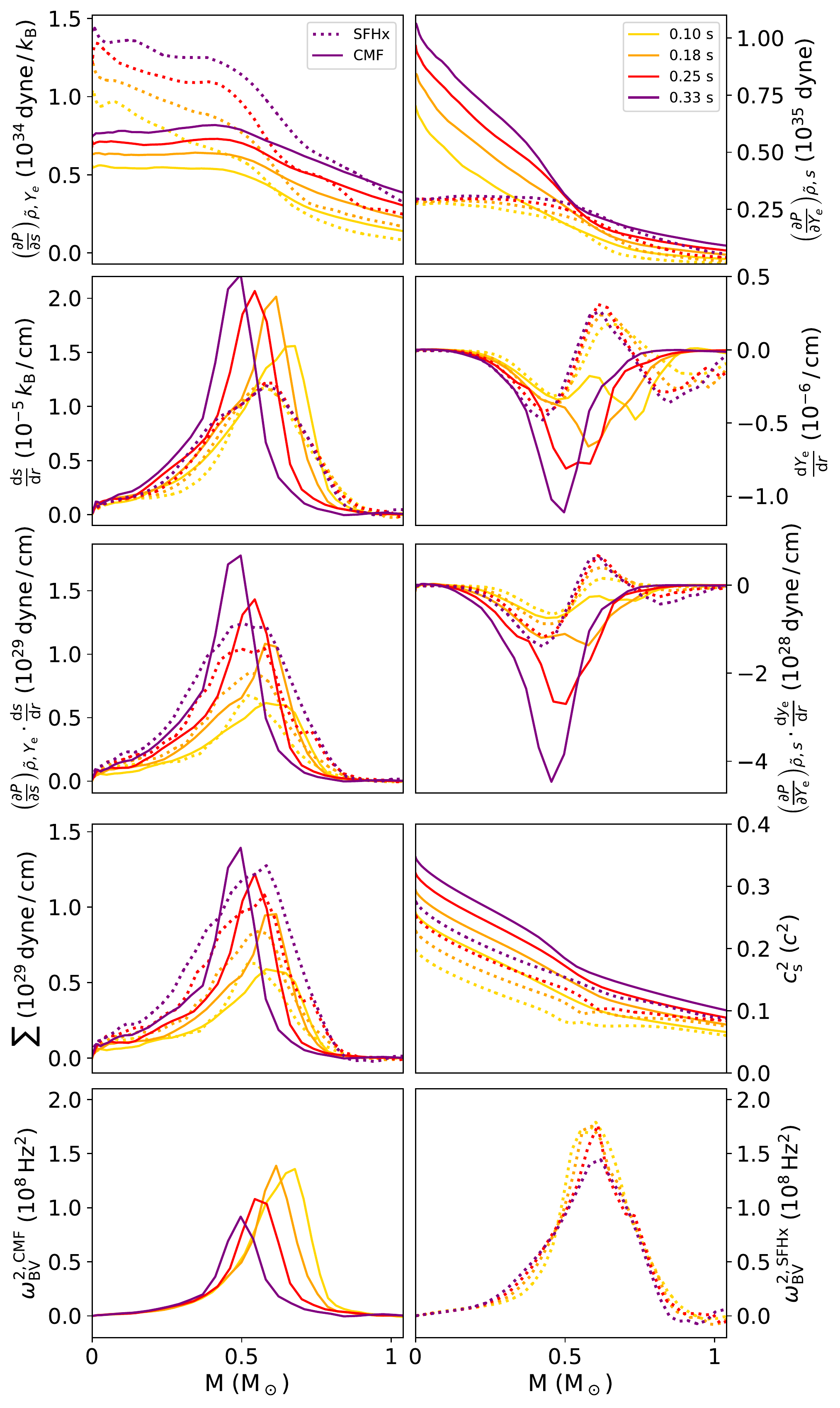}
    \caption{Individual terms for \texttt{z85:CMF} (solid lines) and \texttt{z85:SFHx} (dotted) at four different post-bounce times. See Figure~\ref{fig:z35}.}
    \label{fig:z85}
\end{figure}

The individual terms in Equation~(\ref{eq:bv}) and their products are plotted in Figure~\ref{fig:z85} for the \texttt{z85} models and in Figure~\ref{fig:z35} for the \texttt{z35} models. Each figure shows the results for the \texttt{CMF} models (solid lines) and the \texttt{SFHx} models (dotted) at 4 different times. 


The figures show that 
$\displaystyle{\left(\pdv{ P}{s}\right)_{\!\!\tilde{\rho},Y_\mathrm{e}}}$ and 
$\displaystyle{\dv{s}{r}}$ are quite different for the \texttt{CMF} and \texttt{SFHx} models, but the product
$\displaystyle{\left(\pdv{ P}{s}\right)_{\!\!\tilde{\rho},Y_e} \dv{s}{r}}$ is rather similar.  The electron fraction gradient becomes steeper in the \texttt{CMF} models, which pushes the $\Sigma$-term in square brackets in Equation~(\ref{eq:bv}),
$$\Sigma=\left(\pdv{ P}{s}\right)_{\!\!\tilde{\rho},Y_e}\dv{s}{r}+\left(\pdv{P}{Y_e}\right)_{\!\tilde{\rho},s}\dv{Y_e}{r}$$
to slightly lower values, compared to \texttt{z85:SFHx} where the radial profile of $\left(\pdv{P}{Y_e}\right)_{\!\tilde{\rho},s}\dv{Y_e}{r}$ is flatter and largely positive. The effect is less pronounced for the more massive progenitor \texttt{z85:CMF}; the $\Sigma$-term here is dominated by a larger entropy gradient. For both progenitors, \texttt{z85} and \texttt{z35}, the sound speed in the region below the PCS convection zone is increasingly higher in the \texttt{CMF} models, which systematically pushes $\omega_\mathrm{BV}$ to lower values. In the case of model \texttt{z85}, this effect explains the difference in the Brunt-V\"ais\"al\"a frequency, as the $\Sigma$-term is similar for the
\texttt{SFHx} and \texttt{CMF} EoS.

\section{Radially resolved GW signals}
\begin{figure}[H]
\centering \includegraphics[width=1\columnwidth]{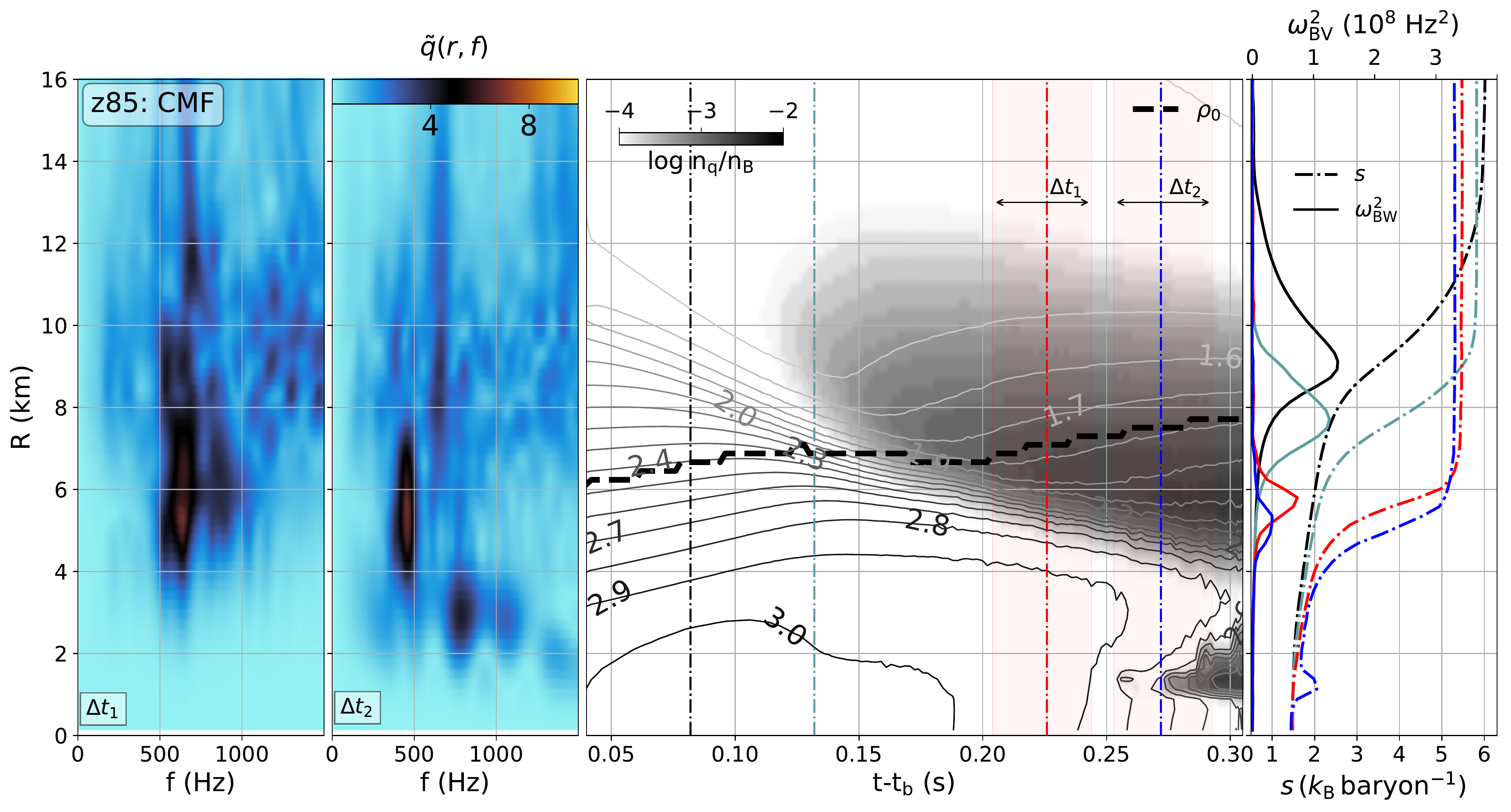}
    \caption{Quadrupolar perturbations, quark fraction, adiabatic index, Brunt-V\"ais\"al\"a frequency and spherically averaged entropy for model
    \texttt{z85:CMF}; see detailed description in text.}
    \label{fig:rgwz85CMF}
\end{figure}

\begin{figure}[H]
\centering \includegraphics[width=1\columnwidth]{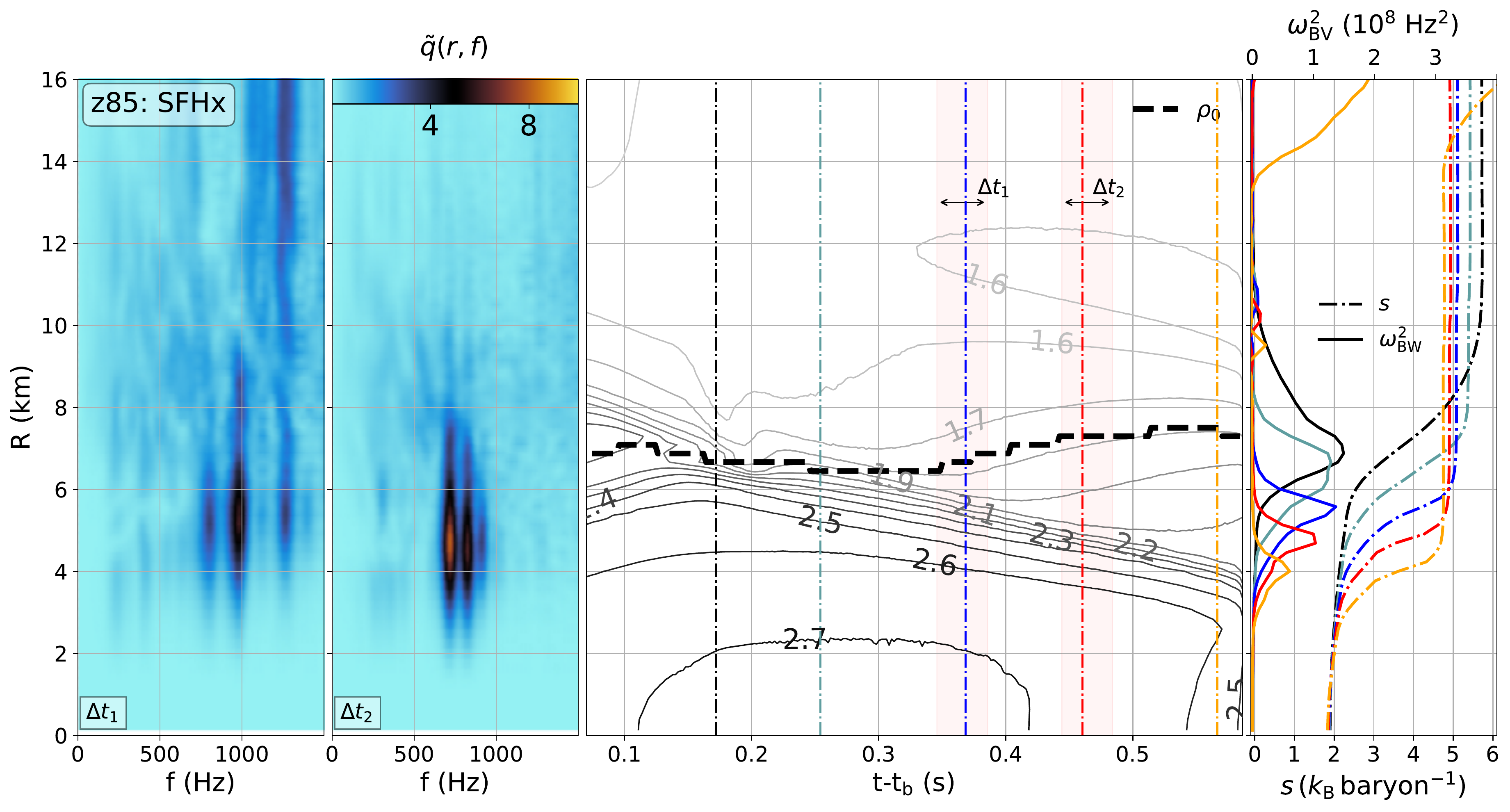}
    \caption{Quadrupolar perturbations, adiabatic index, Brunt-V\"ais\"al\"a frequency and spherically averaged entropy for model \texttt{z85:SFHx}; see detailed description in text. Note that the \texttt{SFHx} EoS is purely hadronic.}
    \label{fig:rgwz85SFHx}
\end{figure}

\begin{figure}[H]
\centering \includegraphics[width=1\columnwidth]{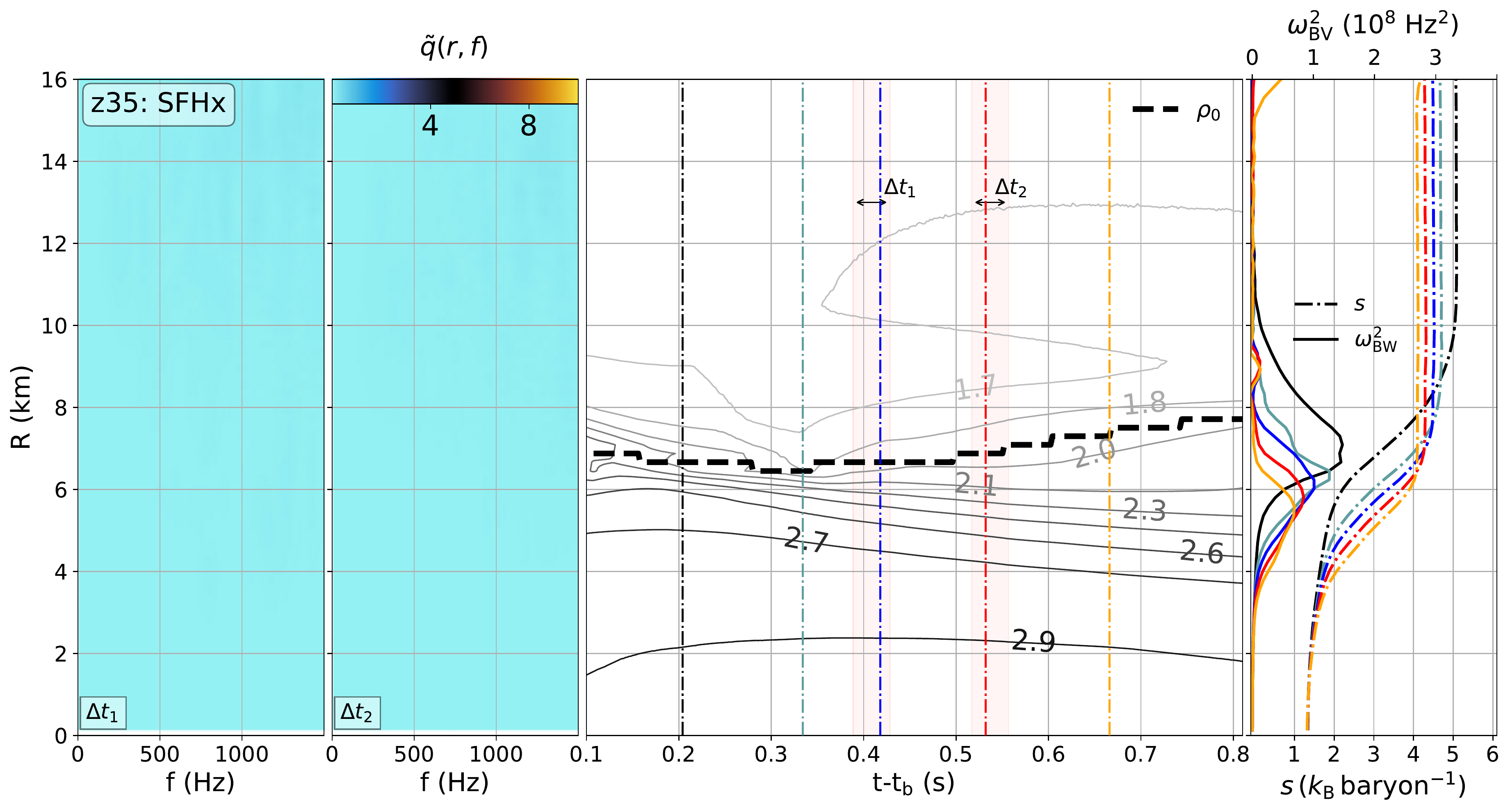}
    \caption{Analogously to Figure~\ref{fig:rgwz85SFHx} but for model \texttt{z35:SFHx} where no signal is observed.}
    \label{fig:rgwz35SFHx}
\end{figure}

For completeness, we show plots analogously to Figure~2 in the main text. Figures~\ref{fig:rgwz85CMF}, \ref{fig:rgwz85SFHx}, and \ref{fig:rgwz35SFHx}
present the results for model \texttt{z85:CMF}, \texttt{z85:SFHx}, and \texttt{z35:SFHx} respectively.
The figures show the radius- and frequency-dependent amplitude $\tilde{q}(r,f)$ of quadrupolar perturbations (first two panels from the left) for two different time intervals $\Delta t_1$ and $\Delta t_2$, which are indicated in the third panel by black and red dash-dotted vertical lines.  The third panels also show the quark fraction (background color; only present in Figure~\ref{fig:rgwz85CMF}) and the adiabatic index (contour lines). The fourth panel shows the squared Brunt-V\"ais\"al\"a frequency and the spherically averaged specific entropy per baryon. 

